\documentclass[twocolumn]{aastex631}
\usepackage{CJK}
\usepackage{amsfonts}
\usepackage{amsmath}
\usepackage{mathrsfs}
\usepackage{epsfig}
\usepackage{subfigure}

\def\nuflub8{\phi^\nu_B}
\def\nuflube7{\phi^\nu_{Be}}

\def\m{\,\mathrm{m}}

\def\s{\,\mathrm{s}}
\def\p{\,\mathrm{p}}

\def\C{\,\mathrm{C}}
\def\N{\,\mathrm{N}}
\def\O{\,\mathrm{O}}

\def\B{\,\mathrm{B}}
\def\A{\,\mathrm{A}}

\def\Rbr{\,R_{\mathrm{br}}}

\def\m{\,\mathrm{m}}

\def\s{\,\mathrm{s}}
\def\p{\,\mathrm{p}}

\def\C{\,\mathrm{C}}
\def\N{\,\mathrm{N}}
\def\O{\,\mathrm{O}}

\def\B{\,\mathrm{B}}
\def\A{\,\mathrm{A}}

\def\Rbr{\,R_{\mathrm{br}}}

\def\R{\mathcal{R}}

\def\s{\,\mathrm{s}}

\newcolumntype{p}{D{,}{\pm}{-1}}

\def\c{\,\mathrm{c}}

\def\g{\,\mathrm{g}}

\shorttitle{Study of the Spectral Hardening @ 100-1000 GV}
\shortauthors{Niu}

\begin{document}
\begin{CJK*}{UTF8}{gbsn}

\title{Quantitative study of the hardening in the Alpha Magnetic Spectrometer nuclei spectra at a few hundred GV}

\correspondingauthor{Jia-Shu Niu}
\email{jsniu@sxu.edu.cn}

\author[0000-0001-5232-9500]{Jia-Shu Niu (牛家树)}
\affil{Institute of Theoretical Physics, Shanxi University, Taiyuan 030006, China}
\affil{State Key Laboratory of Quantum Optics and Quantum Optics Devices, Shanxi University, Taiyuan 030006, China}

\author{Jing Liu}
\affil{Institute of Theoretical Physics, Shanxi University, Taiyuan 030006, China}


\begin{abstract}
The most significant feature in the cosmic-ray (CR) nuclei spectra is the spectral hardening at a few hundred GV. It is important to know whether the hardening of different nuclei species is the same or not for constructing CR sources and propagation models. In this work, we collect the recently released AMS-02 CR nuclei spectra of primary species (proton, helium, carbon, oxygen, neon, magnesium, silicon, and iron), secondary species (lithium, beryllium, boron, and fluorine), and hybrid species (nitrogen, sodium, and aluminum) and study the break positions and the spectral index differences (less and greater than the break rigidity) of the spectral hardening quantitatively. The results show us that the CR nuclei spectral hardening at a few hundred GV has hybrid origins. In detail, the dominating factors of the spectral hardening for primary and secondary CR nuclei species are different: the former comes from the superposition of different kinds of CR sources, while the latter comes from the propagation process. Both of these factors influence all kinds of CR nuclei spectra, just with different weights.
\end{abstract}

\section{Introduction}

The space station experiment Alpha Magnetic Spectrometer (AMS-02) improves the measurement precision of the cosmic-ray (CR) fluxes by an order of the systematics \citep{AMS2013} and deepens our understanding of CRs. Based on the precision data observed by these excellent experiments represented by AMS-02, CR physics has entered a precision-driven era. More and more fine structures have been observed in CR spectra.

Up to now, AMS-02 has released all the spectra of nuclei species up to the atomic number 14 (silicon) based on its first seven years observation, including the primary CR species: proton, helium (He), carbon (C), oxygen (O), neon (Ne), magnesium (Mg), and silicon (Si) \citep{AMS7years,AMS02_Ne_Mg_Si}; the secondary CR species: lithium (Li), beryllium (Be), boron (B), and fluorine (F) \citep{AMS7years,AMS02_F}; the hybrid CR species: nitrogen (N), sodium (Na), and aluminum (Al) \citep{AMS02_N,AMS02_Na_Al}. In addition, the spectrum of heavy primary species iron (Fe) also has been released \citep{AMS02_Fe}.

From an overall perspective, the downward trends of the secondary CR species are more serious than the primary ones, and the hybrid ones are in the middle of both. This corresponds to the origin of the secondary CR nuclei species which are produced in collisions of primary CR particles with the interstellar medium (ISM).
Most of these CR nuclei species show spectral hardening at a few hundred GV, which is the most significant feature in AMS-02 nuclei spectra.
The degrees of the hardening for different CR nuclei species reflect directly to the origin of the hardening, and then point to the features of CR source and propagation \citep{Niu2021}.
With the accumulation of the CR event, the uncertainties in the spectra (especially in high rigidity regions) becomes smaller and smaller. It is both necessary and possible to carry out quantitative studies on these CR nuclei spectra, which could provide us a global view when we go further into the research of CRs.

Physically speaking, the observed CR spectra are produced by the synthetic effects of the  primary source injection spectra, the propagation process, and the solar modulation; even so, it is helpful to analyze the observed CR spectra directly which are always the starting point for building CR models.
Although such kind of works have been simply performed in the AMS-02 data released papers \citep{AMS7years,AMS02_Ne_Mg_Si,AMS02_Na_Al}, they did not always use the independent break power law formulas for different CR nuclei species (such as using one group of parameters to fit the spectral of He, C, and O, and using another group of parameters to fit Ne, Mg, and Si). It would cover the differences between the CR species in one group.
As a result, an independent fittings to each of the CR nuclei species via a uniform method could not only provide us a detailed quantitative comparison between these species, but also give us a global view for guiding the improvements of current CR models.\footnote{Our previous work \citep{Niu2021} performed a similar research based on an old data set from AMS-02, which showed large correlations of systematic errors. An updated data set will give us more reasonable and complete results.}

In the following, we first introduce the methods in Section 2; the results are shown in Section 3; the discussions are presented in Section 4.

\section{Materials and Methods}

Because the spectral hardening happens at a few hundred GV, the data points whose rigidity less than 45 GV are discarded in this work. In such case, we can avoid to handle the solar modulation and fit these CR nuclei spectra using a break power law directly. A break in 100-1000 GV is used to describe the position of the spectral hardening in each of the CR nuclei species. 

The following formula is used to describe each of the AMS-02 nuclei spectra (including the primary CR species: proton, He, C, O, Ne, Mg, Si, and Fe; the secondary CR species: Li, Be, B, and F; the hybrid CR species: N, Na, Al) when the rigidity is greater than 45 GV:
\begin{equation}
  F^{\mathrm{i}}(R) =  N^{\mathrm{i}} \times \left\{ \begin{array}{ll}
                                                       \left( \dfrac{R}{\Rbr^{\mathrm{i}} } \right)^{\nu^{\mathrm{i}}_{1}}  & R \leq \Rbr^{\mathrm{i}}\\
                                                       \left( \dfrac{R}{\Rbr^{\mathrm{i}} } \right)^{\nu^{\mathrm{i}}_{2}} & R > \Rbr^{\mathrm{i}} 
\end{array}
\right.,
\label{eq:bpl}
\end{equation}
where $F$ is the flux of CR, $N$ is the normalization constant, and $\nu_{1}$ and $\nu_{2}$ are the spectral indexes less and greater than the break rigidity $\Rbr$, and $\mathrm{i}$ denotes the species of nuclei.  The errors used in our fitting are the quadratic sum of statistical and systematic errors.

The Markov Chain Monte Carlo (MCMC) framework is employed to determine the posterior probability distributions (PDF) and uncertainties of the spectral parameters for different CR nuclei species.\footnote{The {\sc python} module {\tt emcee} \citep{emcee} is employed to perform the MCMC sampling. Some such examples can be referred to \citet{Niu201801,Niu2019,Niu2022} and references therein.}

\section{Results}

\begin{figure*}[htbp]
  \centering
  \includegraphics[width=0.42\textwidth]{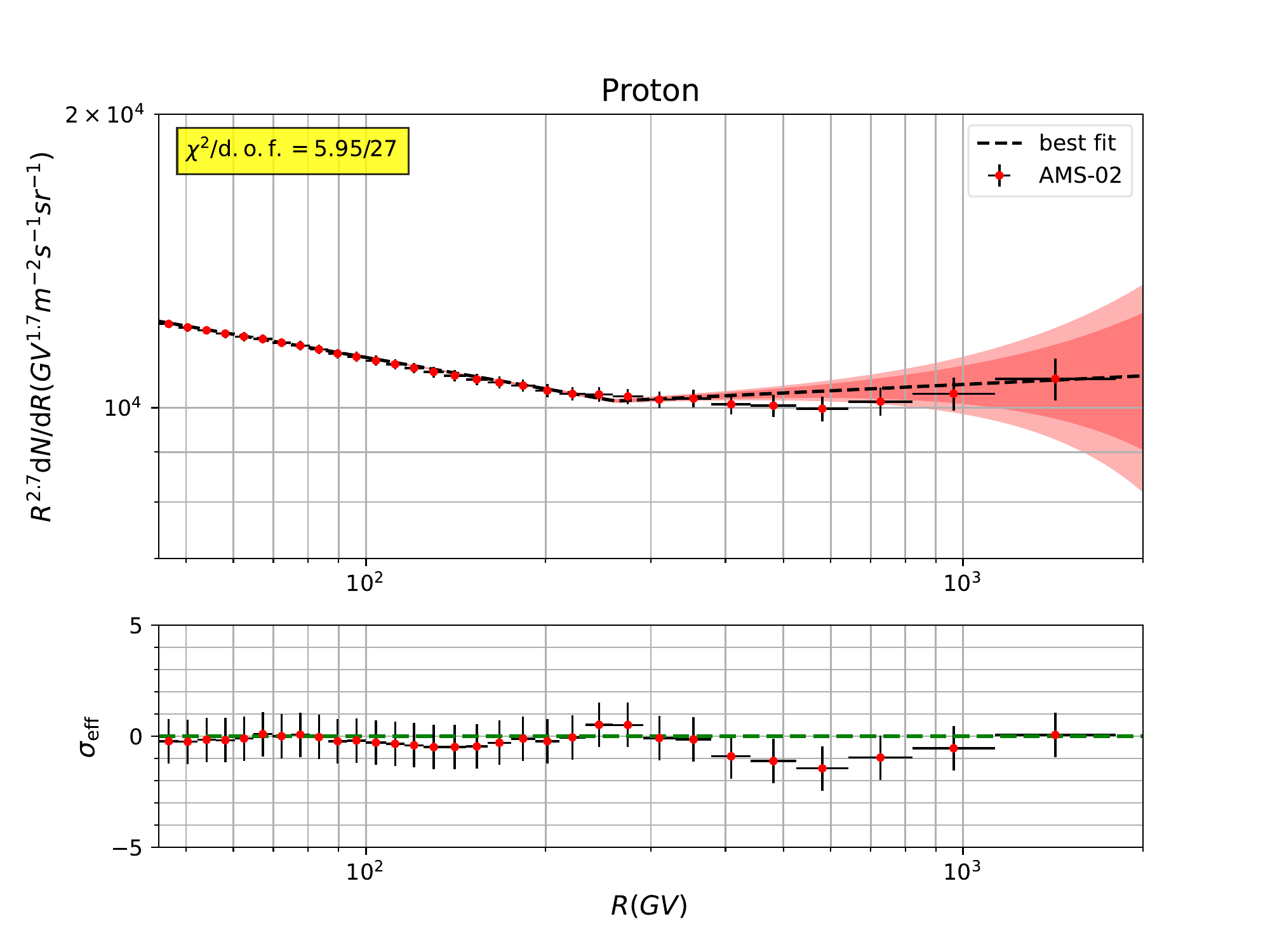}
  \includegraphics[width=0.42\textwidth]{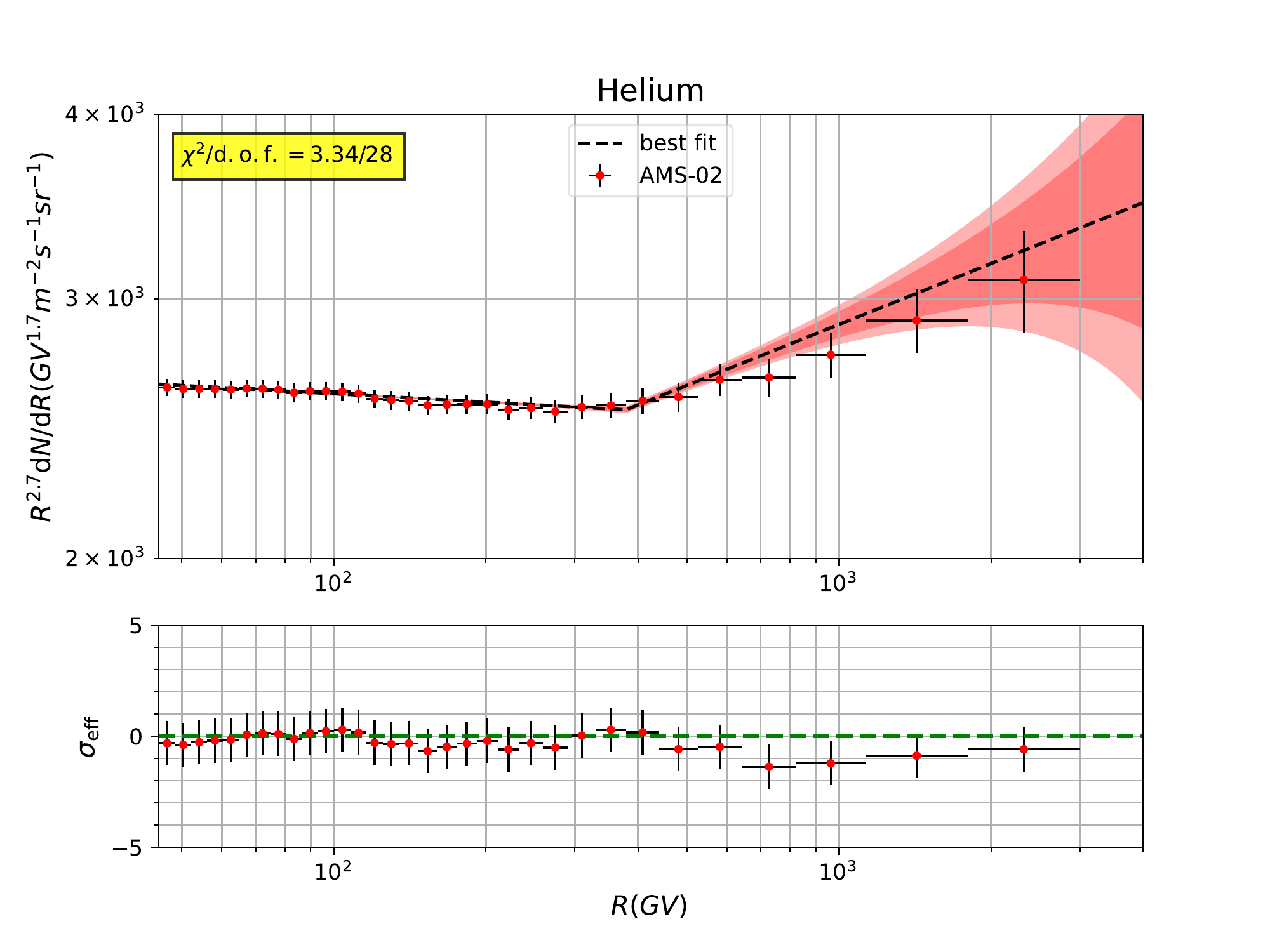}
  \includegraphics[width=0.42\textwidth]{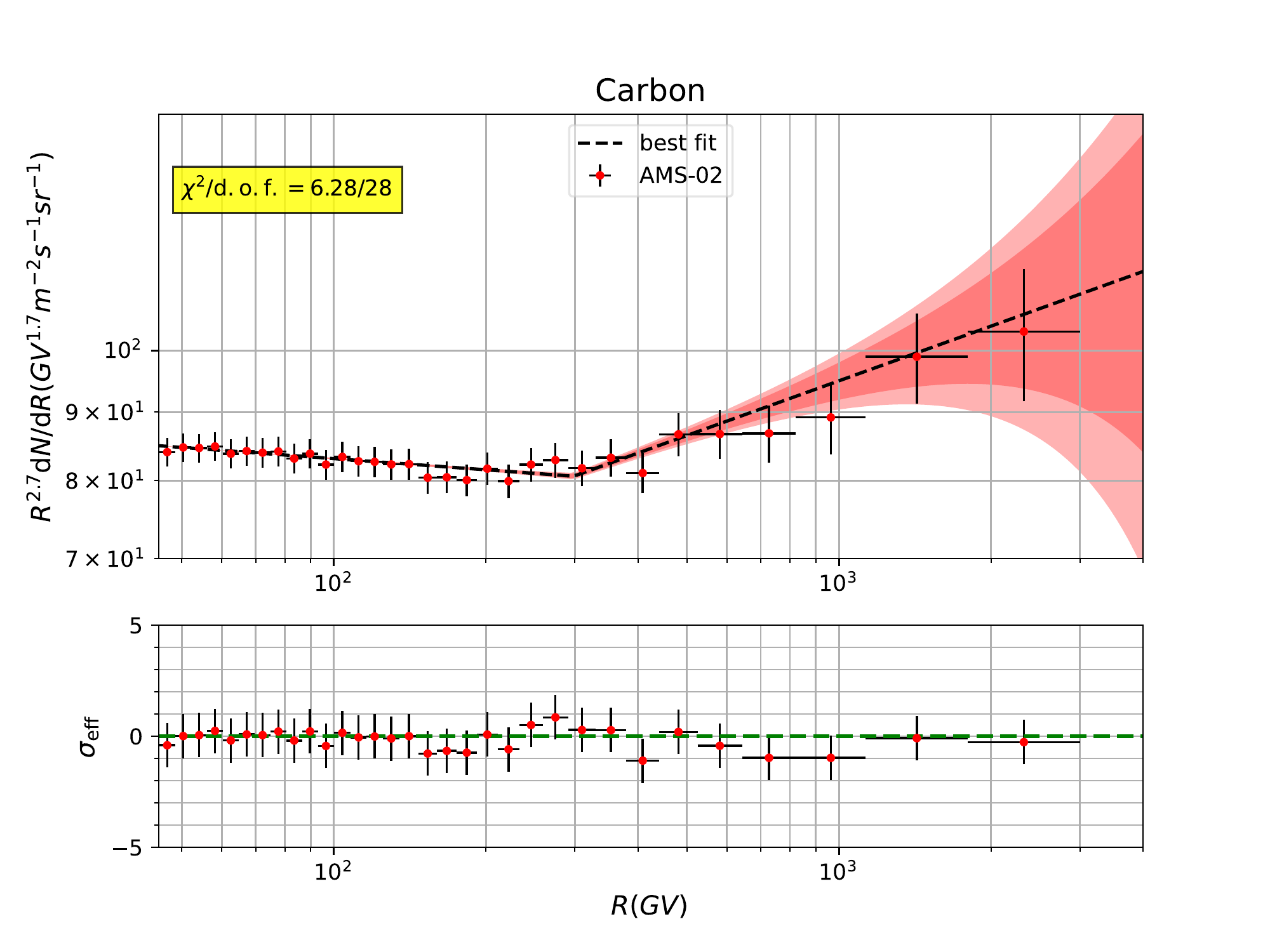}
  \includegraphics[width=0.42\textwidth]{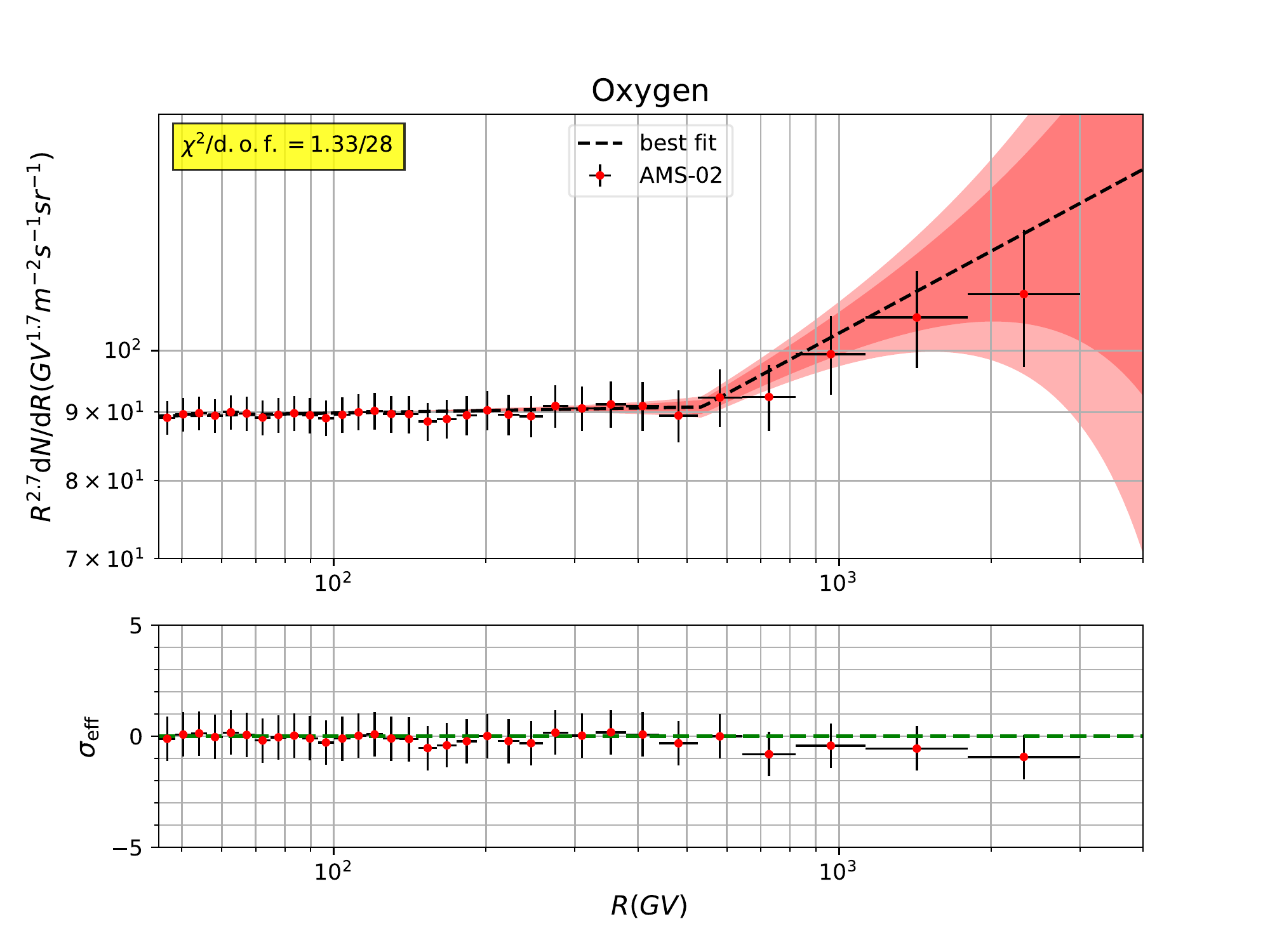}
  \includegraphics[width=0.42\textwidth]{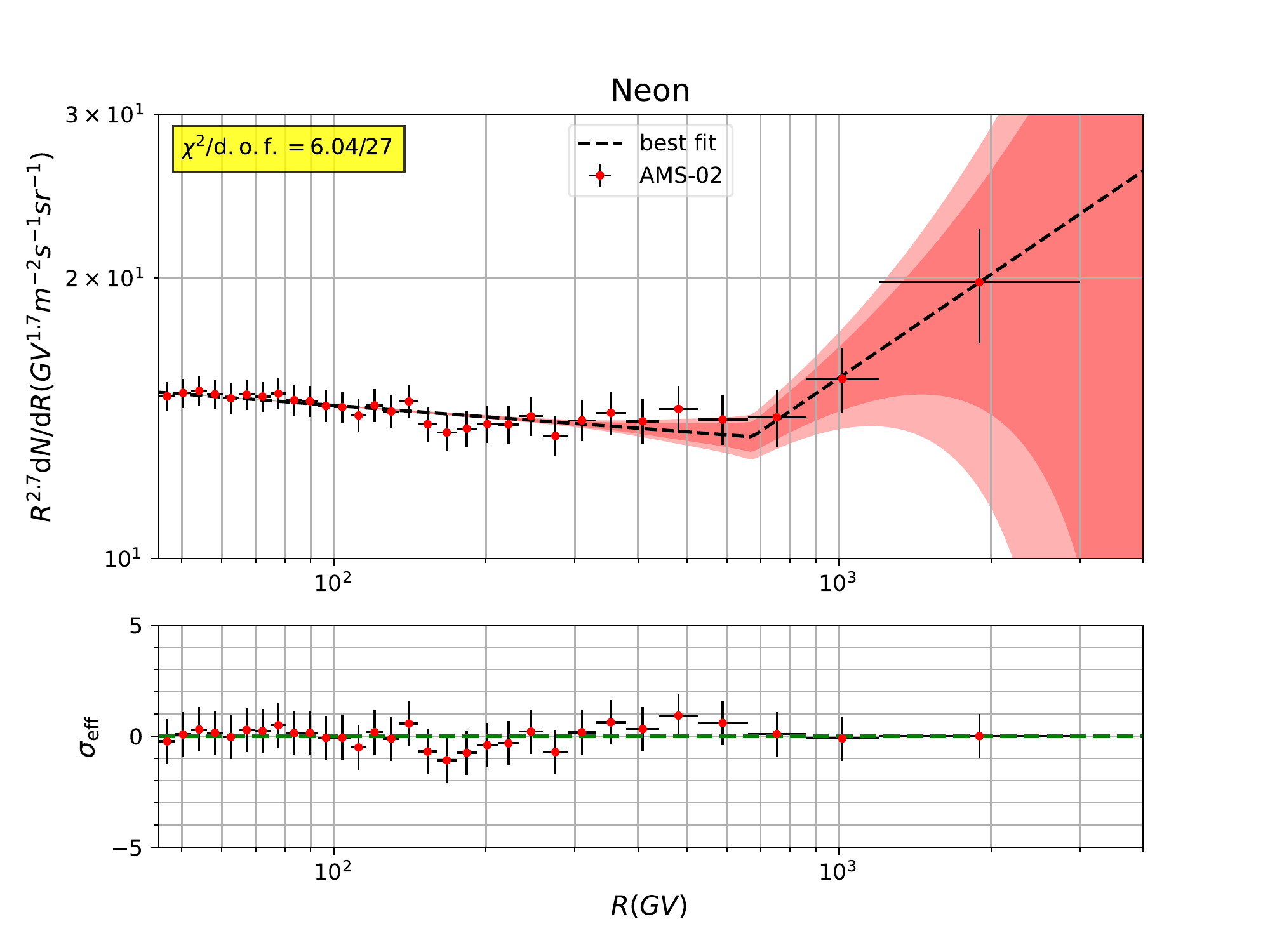}
  \includegraphics[width=0.42\textwidth]{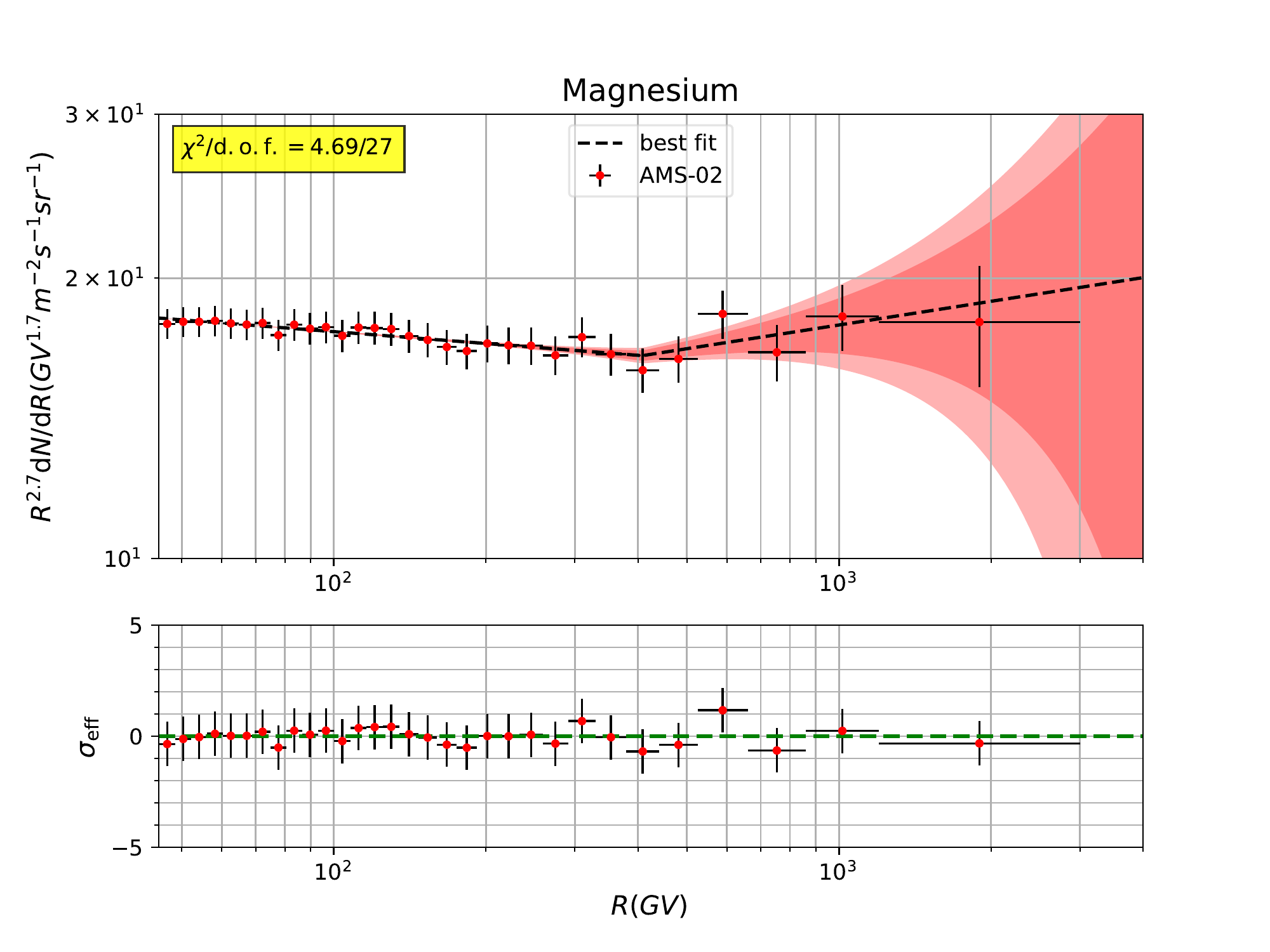}
  \includegraphics[width=0.42\textwidth]{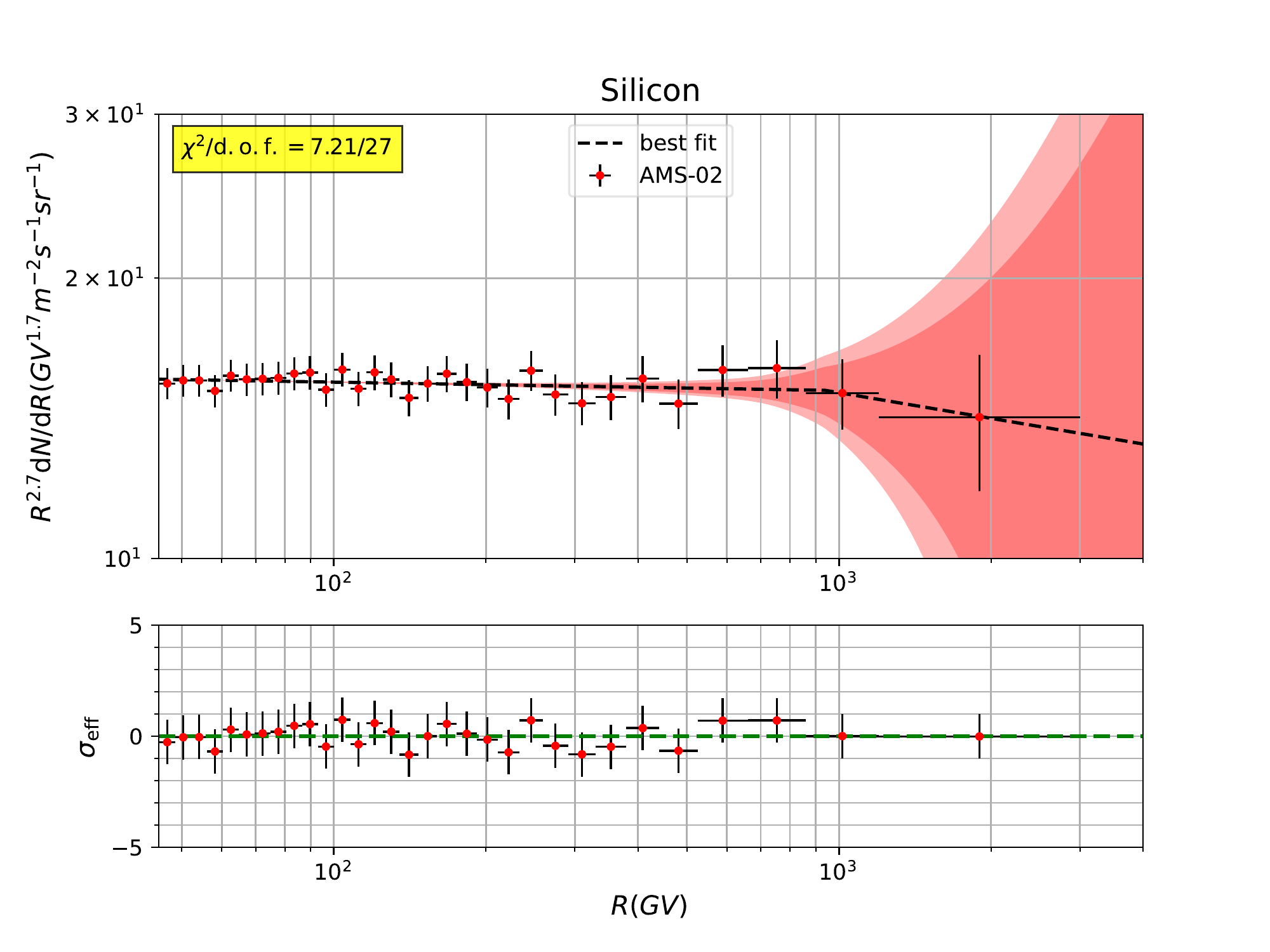}
  \includegraphics[width=0.42\textwidth]{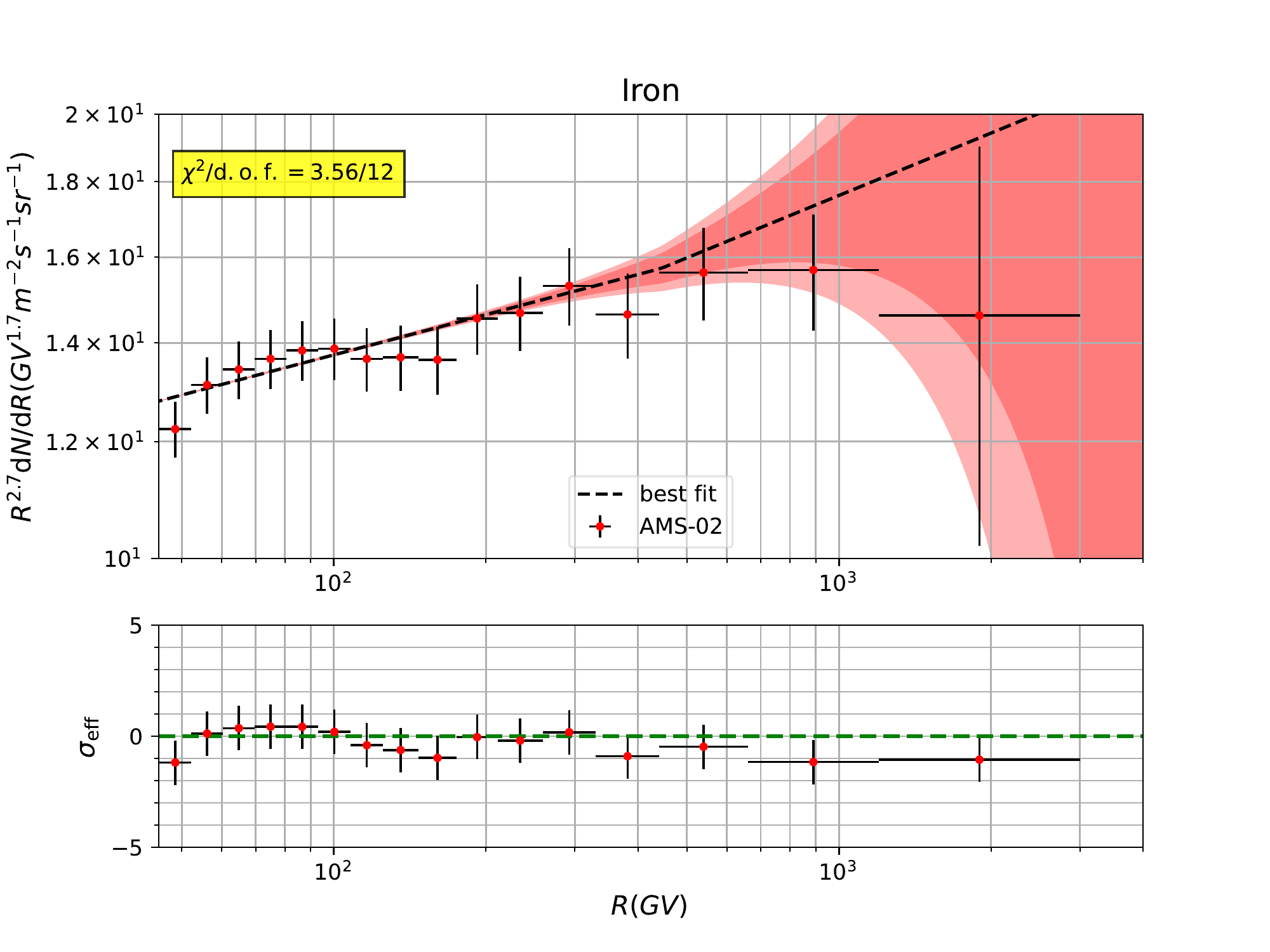}
  \caption{Fitting results and corresponding residuals to the primary CR nuclei spectra (proton, He, C, O, Ne, Mg, Si, and Fe). The 2$\sigma$ (deep red) and 3$\sigma$ (light red) bounds are also shown in the subfigures. The relevant reduced $\chi^2$ of each spectrum is given in the subfigures as well.}
  \label{fig:pri_spectra}
\end{figure*}

\begin{figure*}[htbp]
  \centering
  \includegraphics[width=0.42\textwidth]{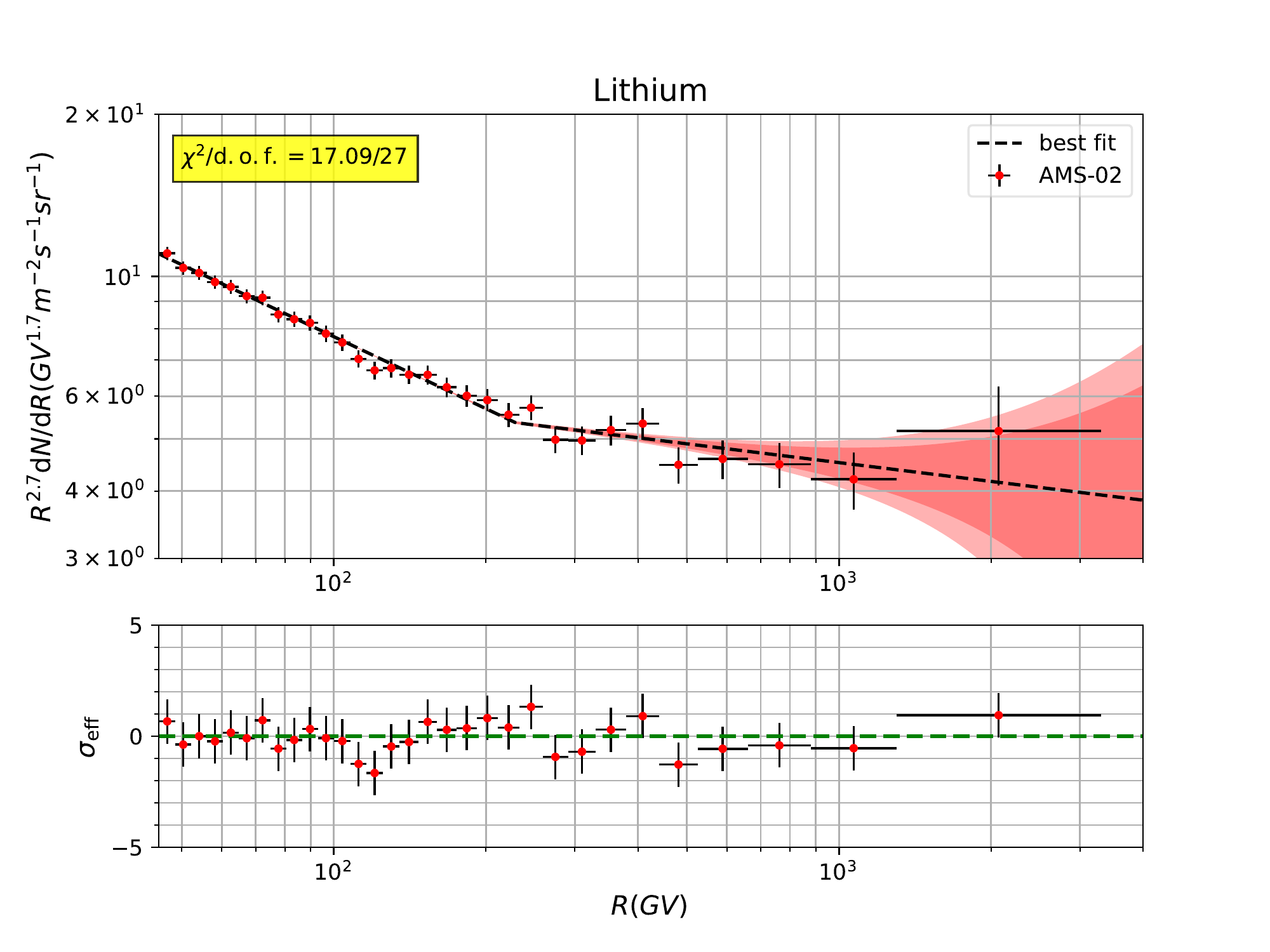}
  \includegraphics[width=0.42\textwidth]{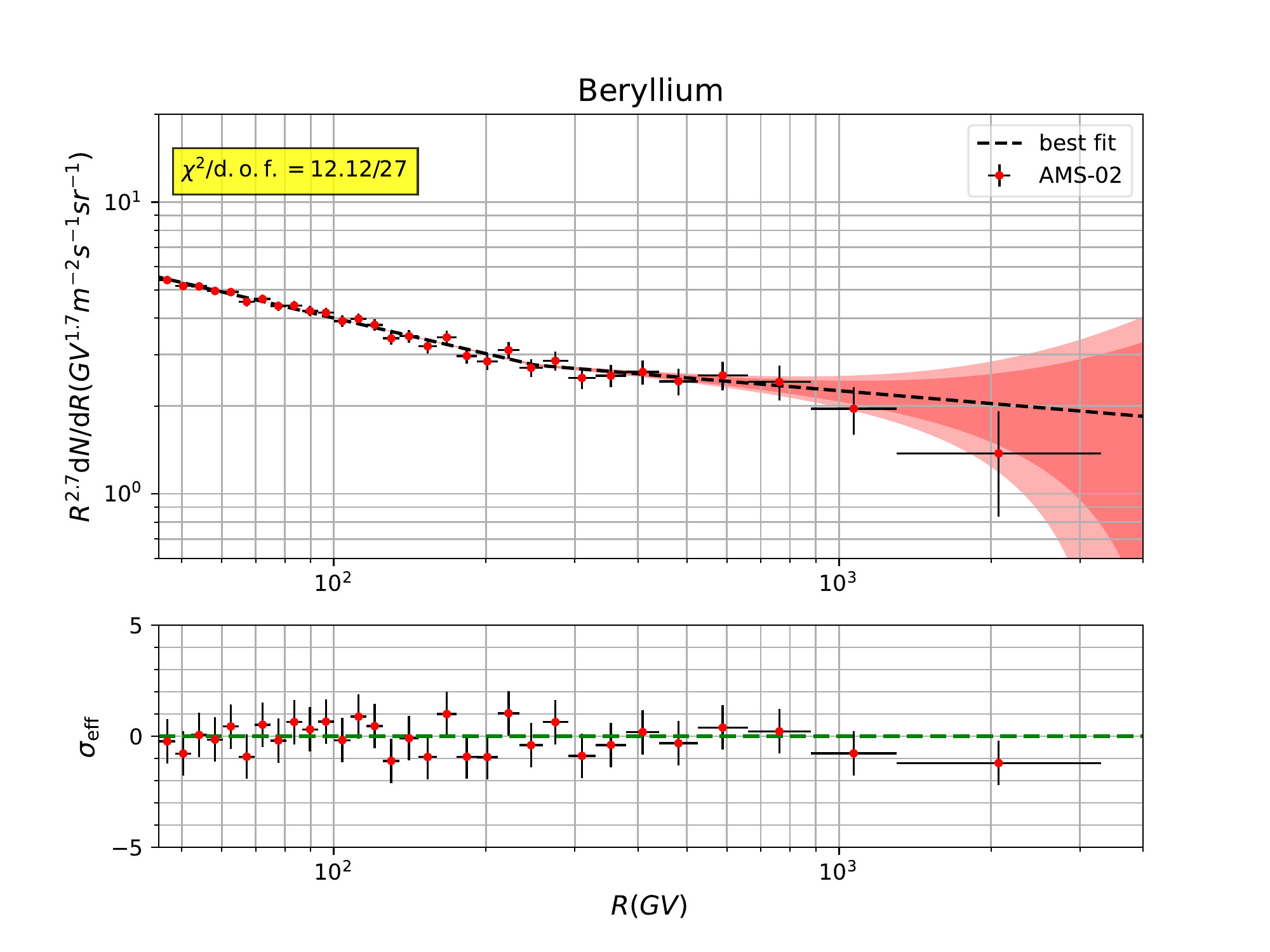}
  \includegraphics[width=0.42\textwidth]{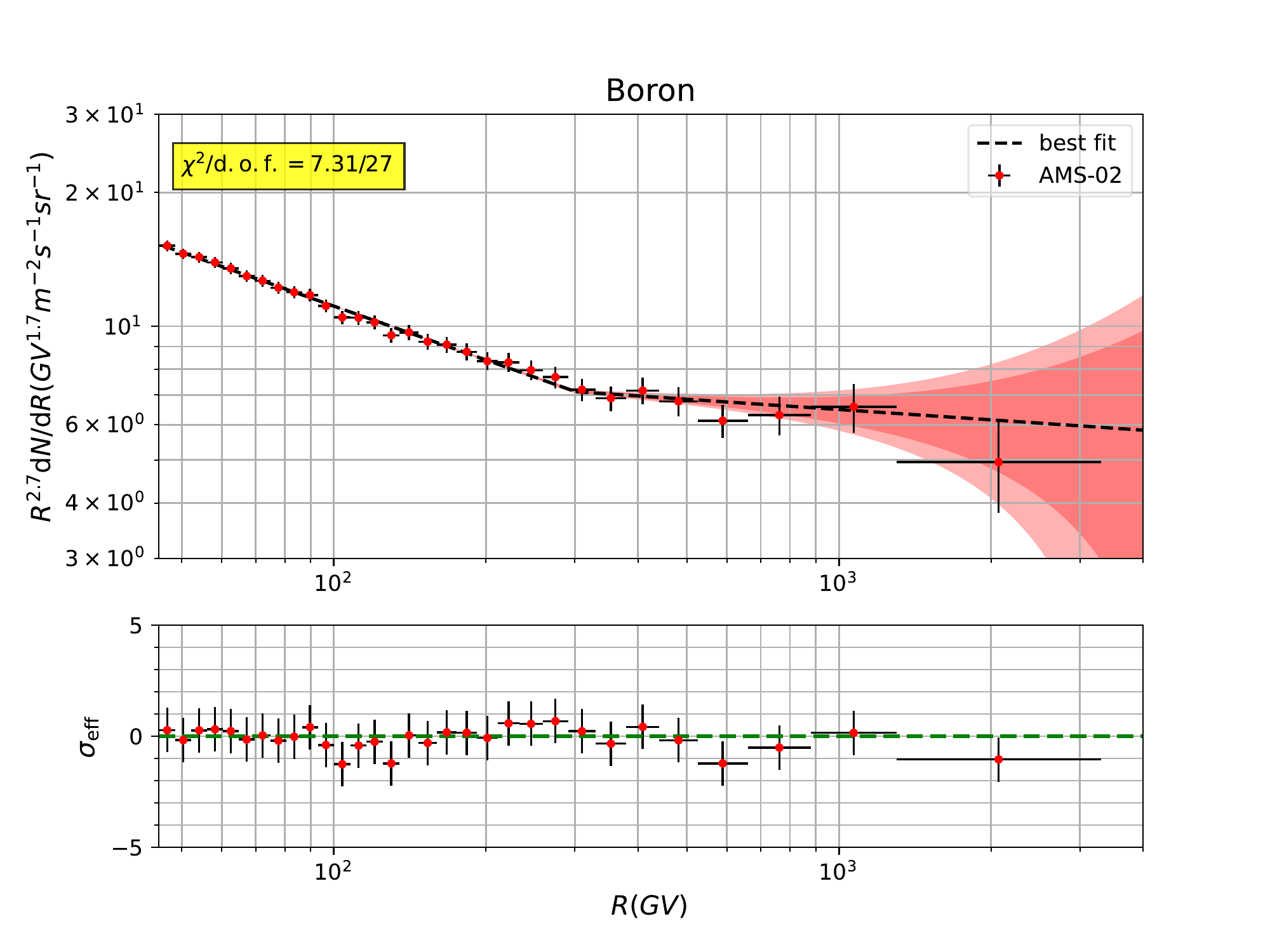}
  \includegraphics[width=0.42\textwidth]{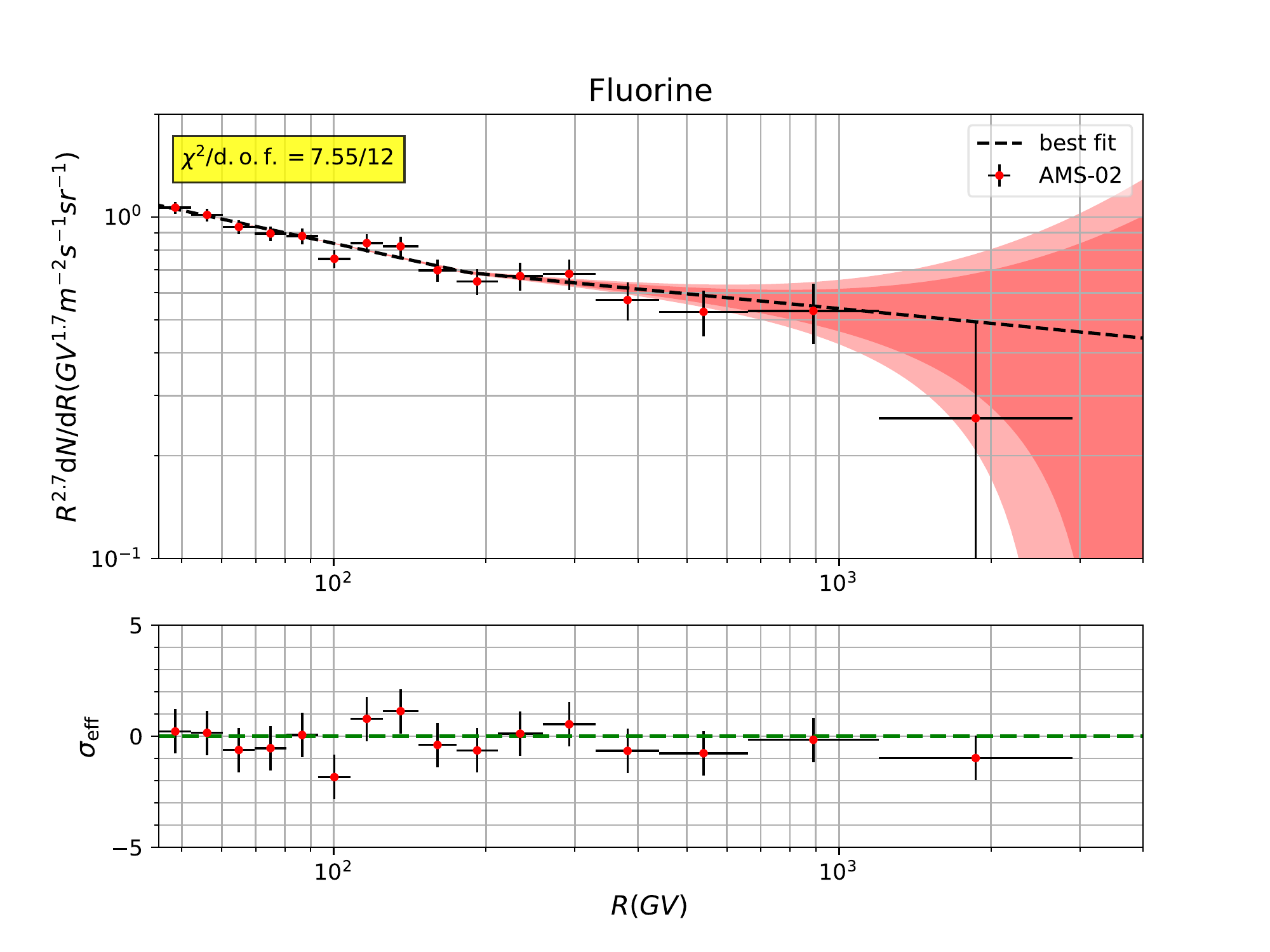}
  \caption{Fitting results and corresponding residuals to the secondary CR nuclei spectra (Li, Be, B, and F). The 2$\sigma$ (deep red) and 3$\sigma$ (light red) bounds are also shown in the subfigures. The relevant reduced $\chi^2$ of each spectrum is given in the subfigures as well.}
  \label{fig:sec_spectra}
\end{figure*}

\begin{figure*}[htbp]
  \centering
  \includegraphics[width=0.42\textwidth]{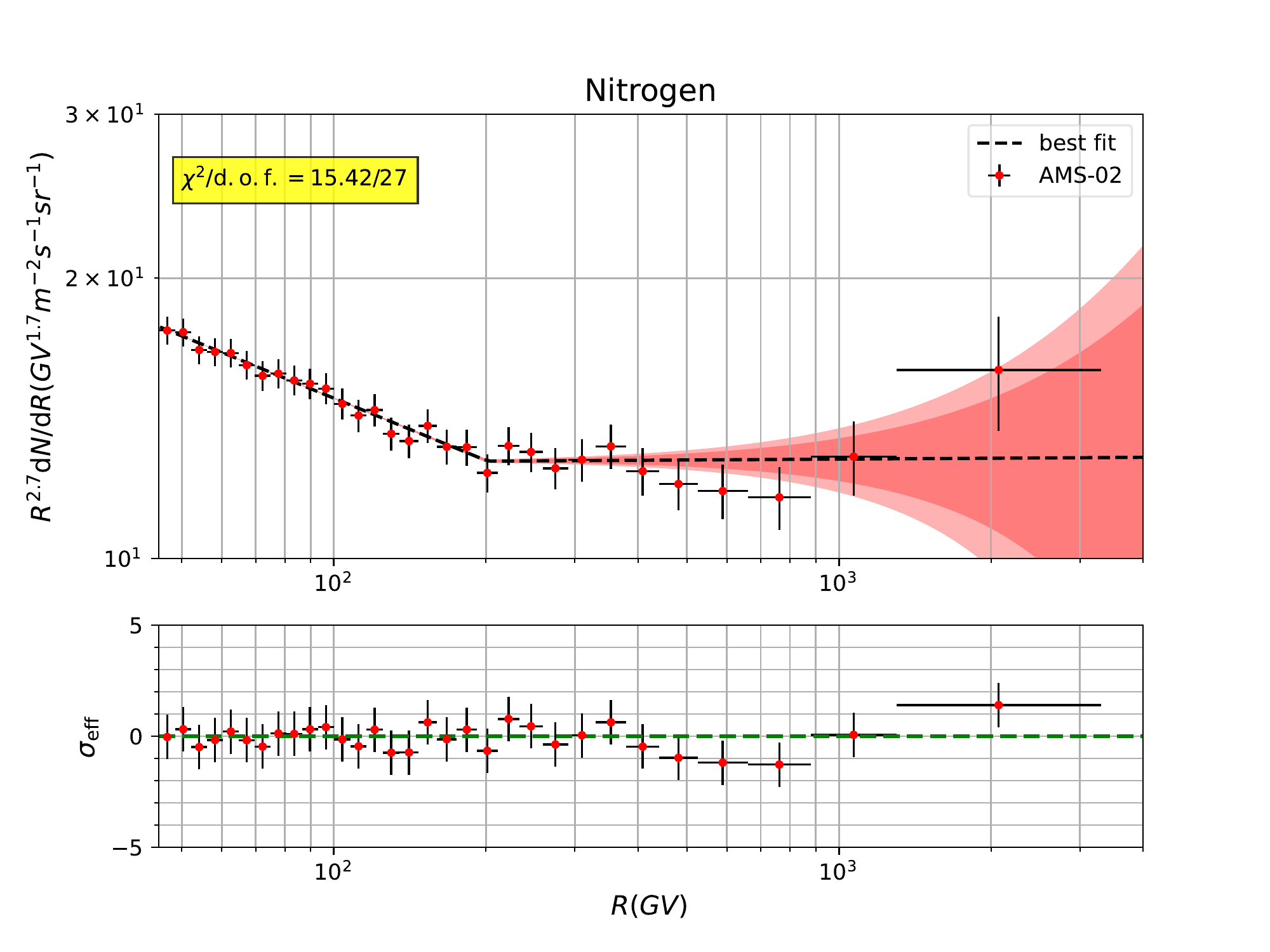}
  \includegraphics[width=0.42\textwidth]{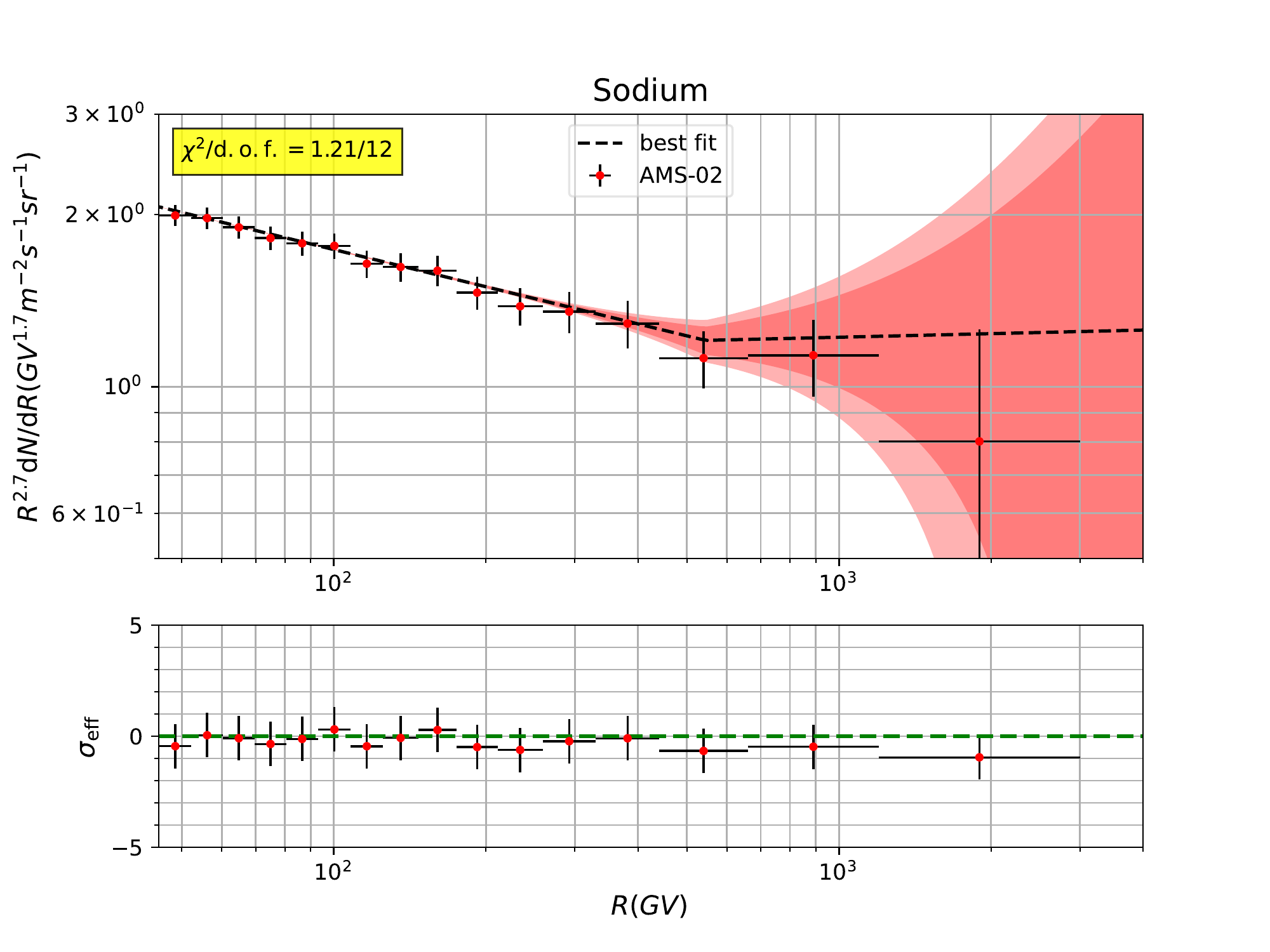}
  \includegraphics[width=0.42\textwidth]{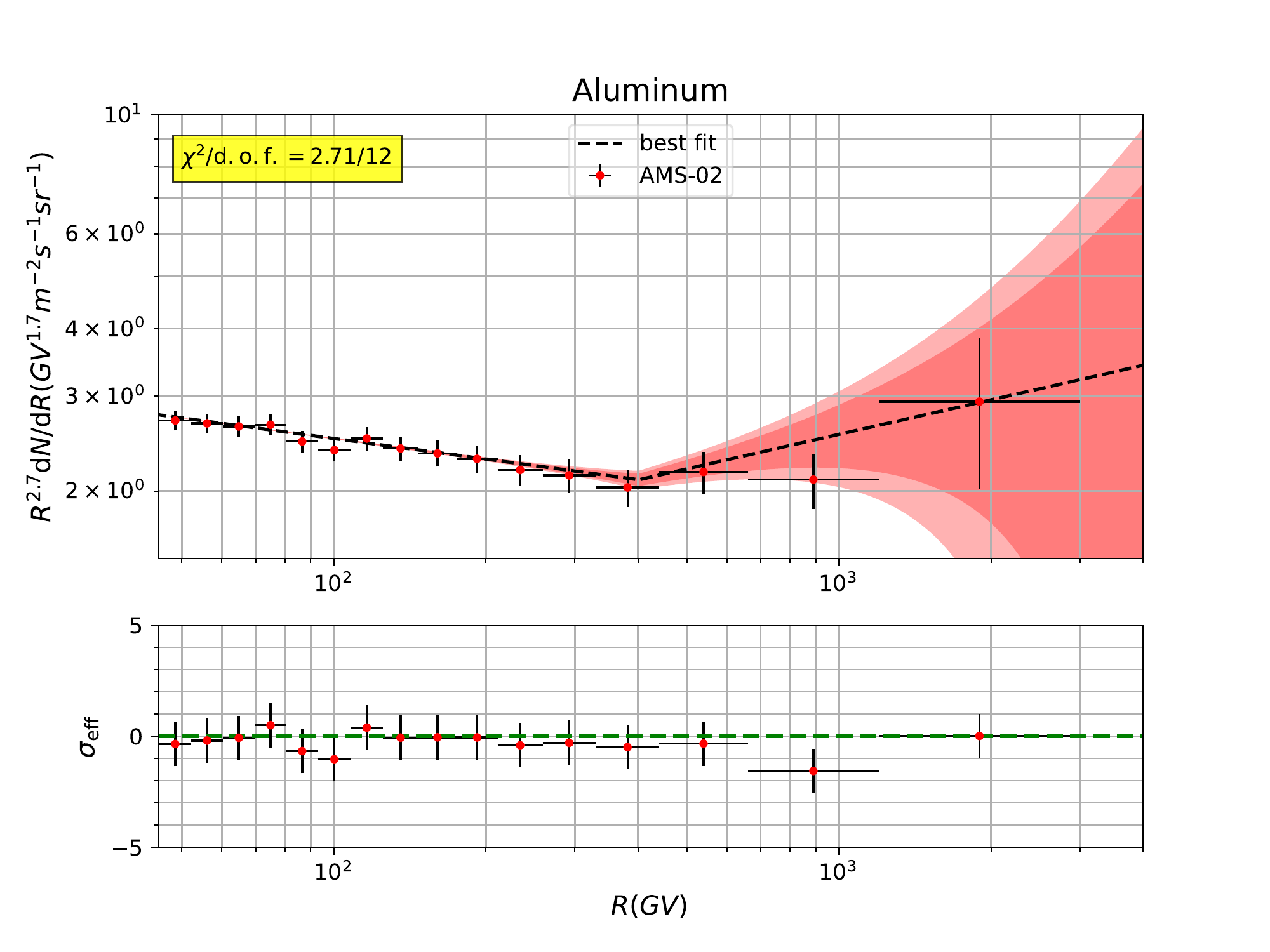}
  \caption{Fitting results and corresponding residuals to the hybrid CR nuclei spectra (N, Na, and Al). The 2$\sigma$ (deep red) and 3$\sigma$ (light red) bounds are also shown in the subfigures. The relevant reduced $\chi^2$ of each spectrum is given in the subfigures as well.}
  \label{fig:hyb_spectra}
\end{figure*}

The best-fit values and the allowed intervals from 5th percentile to 95th percentile of the parameters $\nu_1$, $\nu_2$, $\Rbr$, and $\Delta \nu \equiv \nu_2 - \nu_1$ are listed in Table \ref{tab:spectra_params}, together with the reduced $\chi^2$ of each fitting.\footnote{The information of the parameter $N$ is not listed in the table, which is not important in the subsequent analysis. The PDF of $\Delta \nu \equiv \nu_2 - \nu_1$ is derived from that of $\nu_1$ and $\nu_2$.} The best-fit results and the corresponding residuals of the primary, the secondary, and the hybrid CR species are given in Figure \ref{fig:pri_spectra}, \ref{fig:sec_spectra}, and \ref{fig:hyb_spectra}, respectively.\footnote{Note that in the lower panel of subfigures in Figs. \ref{fig:pri_spectra}, \ref{fig:sec_spectra}, and \ref{fig:hyb_spectra}, the $\sigma_{\mathrm{eff}}$ is defined as
  \begin{equation*}
    \sigma_{\mathrm{eff}} = \frac{f_{\mathrm{obs}} - f_\mathrm{cal}}{\sqrt{\sigma_\mathrm{stat}^{2} + \sigma_\mathrm{syst}^{2}}},
  \end{equation*}
  where $f_\mathrm{obs}$ and $f_\mathrm{cal}$ are the points which come from the observation and model calculation; $\sigma_\mathrm{stat}$ and $\sigma_\mathrm{syst}$ are the statistical and systematic standard deviations of the observed points.}

\begin{table*}
  \caption{The fitting results of the spectral parameters for the different nuclei species. Best-fit values and allowed 5th to 95th percentile intervals (in the square brackets) are listed for each of the parameters. }
\begin{tabular}{c|cccc|c}
\hline\hline
Species  &$\nu_1$  &$\nu_2$  &$\Rbr\ \mathrm{(GV)}$  &$\Delta \nu$  &$\chi^2 / \mathrm{d.o.f}$    \\
\hline
  proton          &-2.808\ [-2.814, -2.796]  &-2.671\ [-2.693, -2.613]  &259\ [244, 348]  &0.137\ [0.112,0.192]  & 5.95/27 = 0.22 \\
  Helium          &-2.719\ [-2.727, -2.709]  &-2.570\ [-2.602, -2.501]  &367\ [318, 488]  &0.149\ [0.115,0.215]  & 3.34/28 = 0.12\\
  Carbon          &-2.727\ [-2.746, -2.707]  &-2.559\ [-2.619, -2.479]  &306\ [217, 438]  &0.168\ [0.107,0.245]  & 6.28/28 = 0.22 \\
  Oxygen          &-2.694\ [-2.709, -2.678]  &-2.500\ [-2.599, -2.392]  &529\ [409, 676]  &0.194\ [0.093,0.306]  & 1.33/28 = 0.05 \\
  Neon            &-2.741\ [-2.759, -2.719]  &-2.362\ [-2.568, -2.079]  &660\ [542, 849]  &0.379\ [0.161,0.658]  & 6.04/27 = 0.22 \\
  Magnesium       &-2.742\ [-2.765, -2.721]  &-2.609\ [-2.724, -2.529]  &414\ [323, 464]  &0.133\ [0.011,0.219] & 4.69/27 = 0.17 \\
  Silicon         &-2.709\ [-2.729, -2.690]  &-2.792\ [-3.299, -2.477]  &923\ [873, 994]  &-0.083\ [-0.585,0.230]  & 7.21/27 = 0.27 \\  
  Iron            &-2.614\ [-2.647, -2.573]  &-2.542\ [-2.756, -2.390]  &392\ [315, 544]  &0.072\ [-0.171,0.239]  & 3.56/12 = 0.30 \\  
\hline
  Lithium         &-3.146\ [-3.174, -3.109]  &-2.836\ [-2.905, -2.665]  &216\ [177, 322]  &0.310\ [0.239,0.470]  & 17.09/27 = 0.63 \\
  Beryllium       &-3.102\ [-3.131, -3.066]  &-2.848\ [-2.967, -2.657]  &247\ [195, 415]  &0.254\ [0.122,0.449]  & 12.12/27 = 0.45 \\
  Boron           &-3.103\ [-3.128, -3.075]  &-2.765\ [-2.899, -2.601]  &308\ [225, 436]  &0.338\ [0.201,0.492]  & 7.31/27 = 0.27 \\
  Fluorine        &-3.016\ [-3.091, -2.951]  &-2.844\ [-3.007, -2.712]  &189\ [163, 216]  &0.172\ [-0.289,0.347]  & 7.55/12 = 0.63 \\
\hline
  Nitrogen        &-2.925\ [-2.955, -2.883]  &-2.694\ [-2.753, -2.574]  &199\ [157, 325]  &0.231\ [0.158,0.335]  & 15.42/27 = 0.57 \\
  Sodium          &-2.913\ [-2.958, -2.873]  &-2.657\ [-3.213, -2.372]  &558\ [442, 629]  &0.256\ [-0.293,0.558]  & 1.21/12 = 0.10 \\
  Aluminum        &-2.827\ [-2.870, -2.785]  &-2.487\ [-2.733, -2.268]  &401\ [270, 500]  &0.340\ [0.076,0.574]  & 2.71/12 = 0.23 \\
\hline  
\end{tabular}
\label{tab:spectra_params}
\end{table*}

Generally speaking, the reduced $\chi^2$s of all the CR nuclei species are smaller than 1.0, which indicates the success of the break power law to describe the spectra. But some too small reduced $\chi^{2}$s imply an improper treatment of the data errors. The improvement of the treatment needs additional information about the correlation matrix of systematic errors of AMS-02 data. Some detailed discussions of this topic can be found in \citet{Derome2019,Weinrich202001,Heisig2020}. One should note that the reduced $\chi^{2}$s in Table \ref{tab:spectra_params} do not have the absolute meaning of goodness-of-fit, although they can be compared with each other.

\section{Discussions}

\begin{figure*}
\begin{center}
  \includegraphics[width=0.495\textwidth]{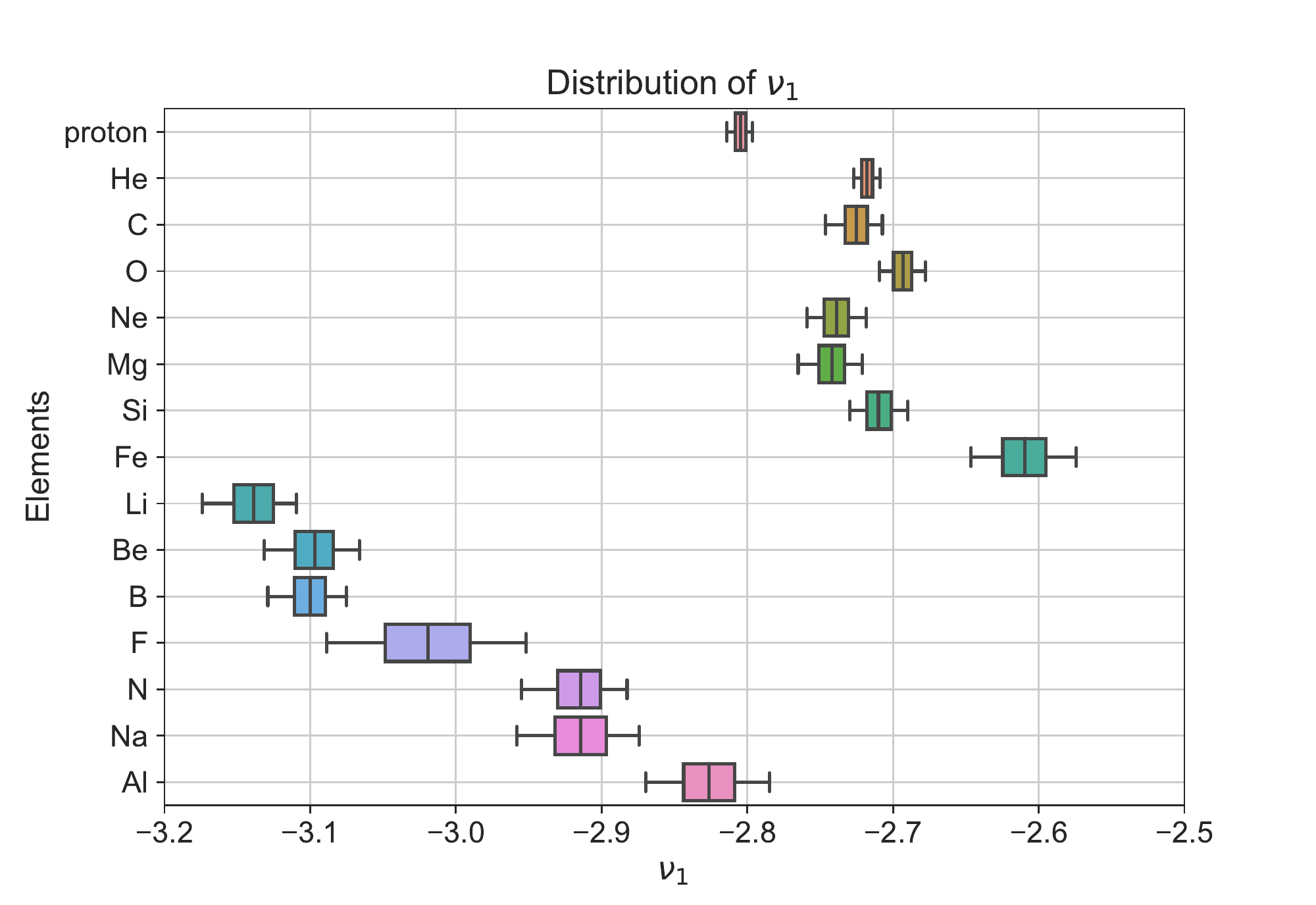}
  \includegraphics[width=0.495\textwidth]{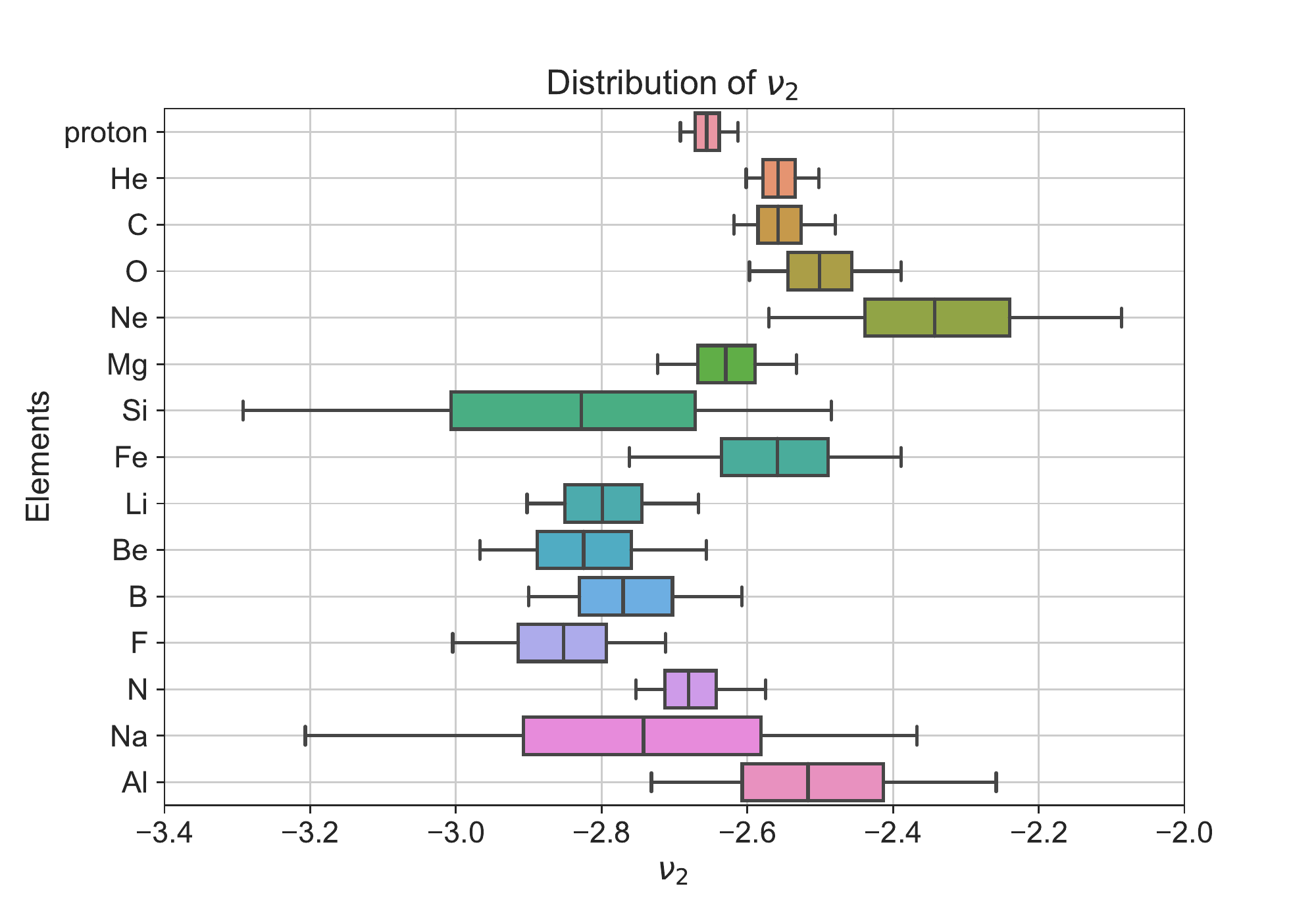}
  \includegraphics[width=0.495\textwidth]{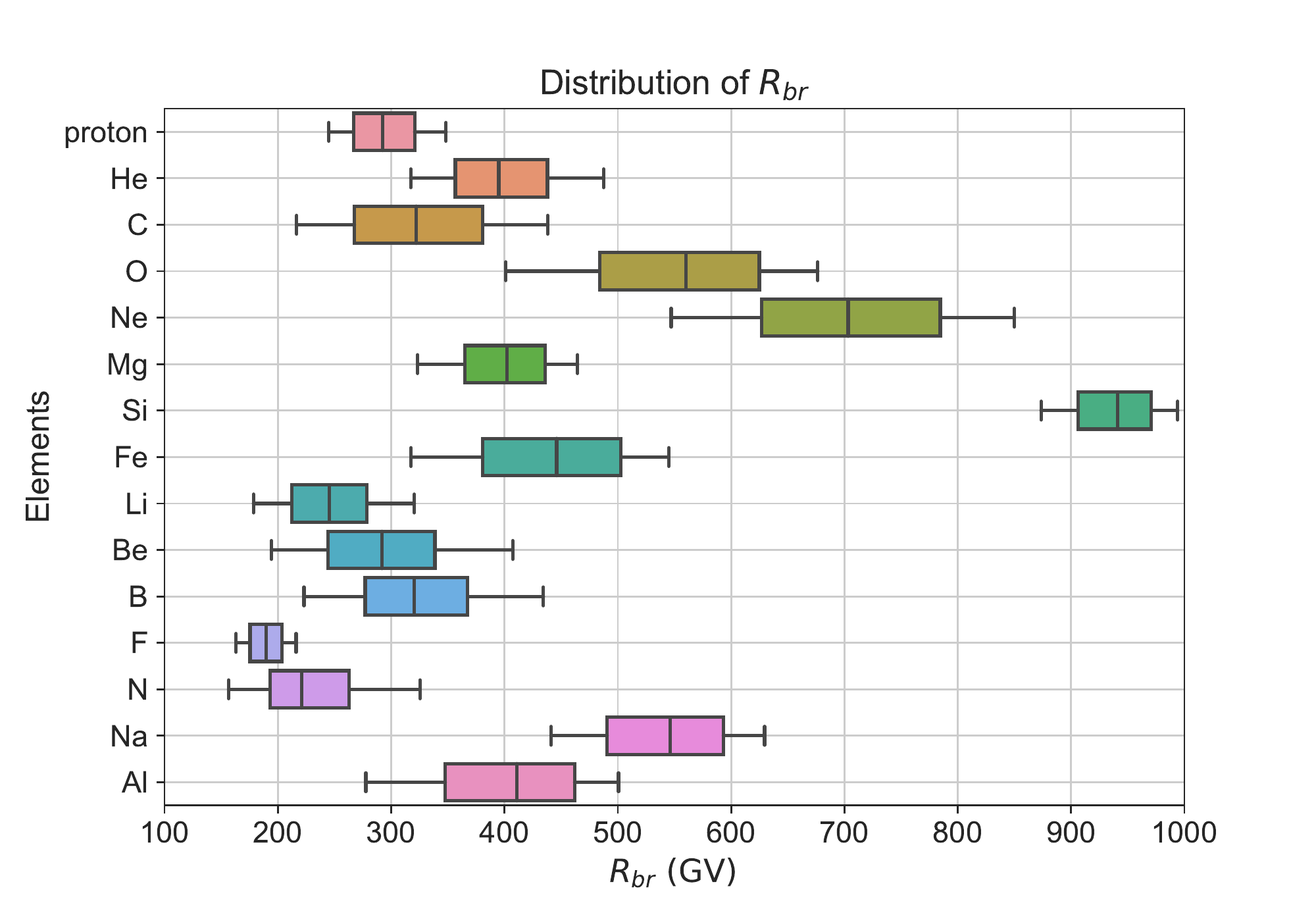}
  \includegraphics[width=0.495\textwidth]{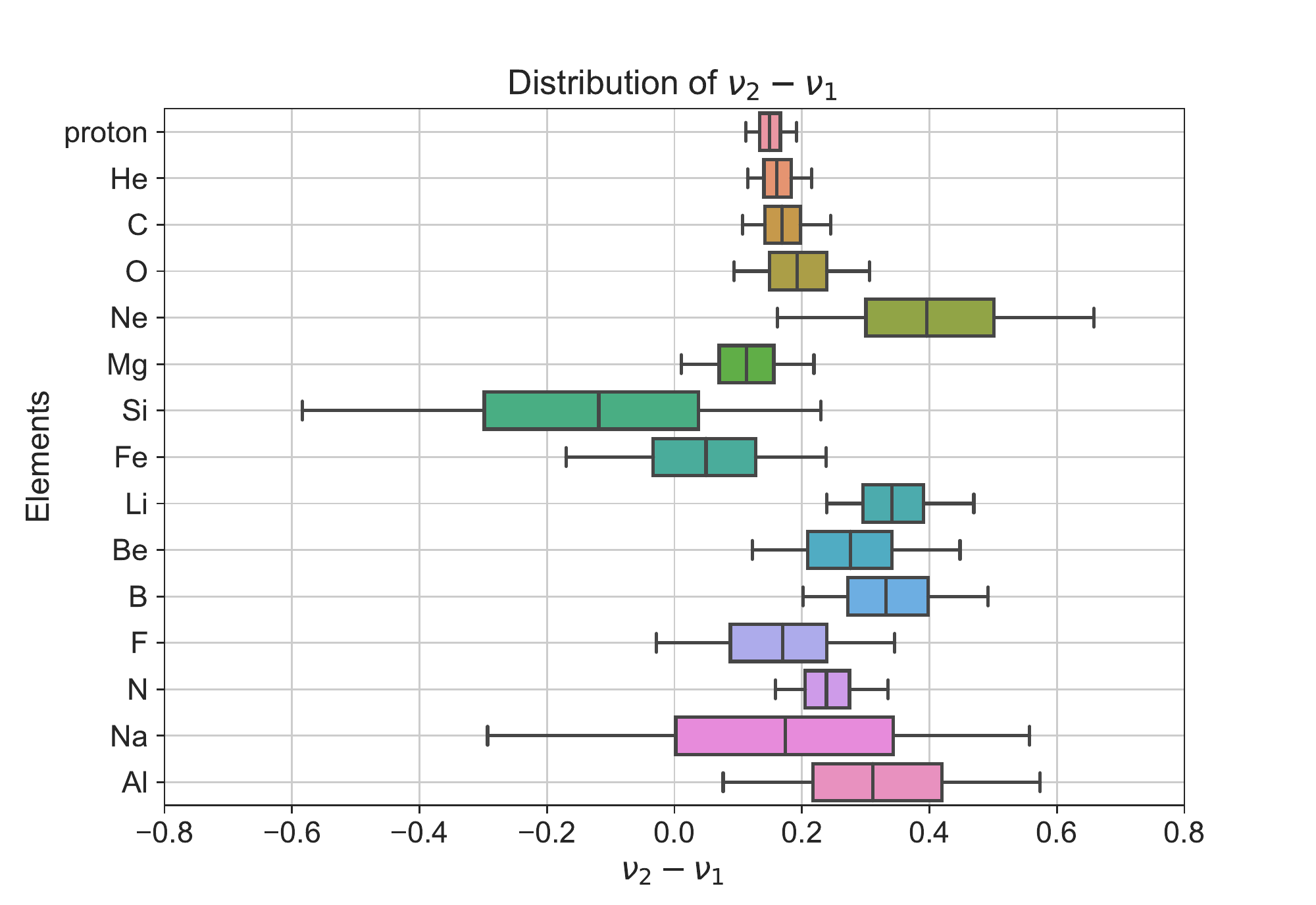}
\end{center}
\caption{Boxplots for $\nu_1$, $\nu_2$, $\Rbr$, and $\nu_{2}-\nu_{1} \equiv \Delta \nu$. The band inside the box shows the median value of the dataset, the box shows the quartiles, and the whiskers extend to show the rest of the distribution which are edged by the 5th percentile and the 95th percentile.}
  \label{fig:spectra_params}
\end{figure*}

In order to get a clear representation and comparison of the fitted parameters of the different nuclei species, the boxplots\footnote{A box plot or boxplot is a method for graphically depicting groups of numerical data through their quartiles. In our configurations, the band inside the box shows the median value of the dataset, the box shows the quartiles, and the whiskers extend to show the rest of the distribution which are edged by the 5th percentile and the 95th percentile.} of the spectral parameters are used to show the distributions of $\nu_1$, $\nu_2$, $\Rbr$, and $\Delta \nu \equiv \nu_2 - \nu_1$ in Figure \ref{fig:spectra_params}.

In the boxplots of $\nu_1$ in Figure \ref{fig:spectra_params}, the secondary species have smallest values, then the hybrid species, and then the primary CR species. It is obvious that the $\nu_1$ values of proton and Fe are significantly different from that of the other primary CR species. The former are always referred to the p/He anomaly, which is generally ascribed to the particle-dependent acceleration mechanisms occurring in galactic CR sources (see for e.g. \citet{Vladimirov2012}). And many specific mechanisms are proposed to interpret this anomaly (see for e.g. \citet{Erlykin2015,Malkov2012,Fisk2012,Ohira2011,Tomassetti2015apjl01}). The later comes from the significantly larger interaction cross sections with the ISM of Fe than those of lighter nuclei species (He, C, O, Ne, Mg, and Si) \citep{AMS02_Fe}.
For the secondary CR nuclei species, the $\nu_1$ value of F is larger than that of others, which indicates that the propagation of heavy CRs (from F to Si) might be different from that of light CRs (from He to O).
For the hybrid CR nuclei species, the $\nu_1$ value of Al is larger than that of N and Na. This is a direct sign of its higher proportion of primary components compared with N and Na (see \citet{AMS02_N,AMS02_Na_Al} for more details).
One should note that a trend is implied in this subfigure: if we consider the $\nu_1$ values in one group of CR nuclei species (primary, secondary or hybrid), they increase with the increasing of atomic number. Whether this trend is just a coincidence, or it comes from an undiscovered mechanism (such as a charge or mass dependent acceleration or propagation) should be tested in future.

In the boxplots of $\nu_2$ in Figure \ref{fig:spectra_params}, the uncertainties are larger than that of $\nu_1$ because of the fewer data points with larger uncertainties in high rigidity region.
  Roughly speaking, the $\nu_2$ values of the primary CR nuclei species are larger than that of the secondary species (except that of Si with quite large uncertainty).
  For hybrid CR nuclei species, N has a $\nu_2$ value with low uncertainty of about 2\%, which is the same as the one for proton and within the uncertainty of the other primary species; the $\nu_2$ uncertainty of Na is about 20\%, which is similar to that of Si (about 18\% - 19\%); the $\nu_2$ of Al has an uncertainty of about 9\% - 10\%, and is similar to the values of the primary ones, which indicates its flux in high rigidity region is dominated by the primary component.


In the boxplots of $\Rbr$ in Figure \ref{fig:spectra_params}, it shows that the break positions are significantly different between some of the CR nuclei species, especially in the case of primary and hybrid species. On the contrary, the break positions of the secondary CR nuclei species are distribute around 200-400 GV, which are a bit more concentrated and indicate they might have a common origin. If the spectral hardening of the secondary CR nuclei species (Li, Be, B, and F) mainly comes from their parents species (C, N, O, Ne, Na, Mg, Al, Si), their break position should have similar distributions.
Considering the heavy secondary CR nuclei species F, it is thought to be produced mostly by the collisions of heavy nuclei (such as Ne, Mg, and Si) with the ISM, but its break position distributes around 200 GV, which is significantly different from its parents species (all of them are larger than 300 GV). This is a definite evidence that the spectral hardening in the secondary CR nuclei species does NOT dominately inherit from its parents species, and the main factor of their hardening comes from propagation (such as in \citet{Blasi2012,Tomassetti2012,Tomassetti2015apjl01,Tomassetti2015prd,Feng2016,Genolini2017,Jin2016CPC,Guo2018cpc,Guo2018prd,Liu2018,Niu2019,Boschini2020apj,Boschini2020apjs,Niu2022}).
Moreover, such diffuse distributions of the break positions of the primary CR nuclei species cannot be dominately reproduced by a uniform acceleration mechanism in CR sources or in propagation process, and the superposition of different kinds of sources (with different spectral index and element abundances) seems to be the only natural explanation (such as in \citet{Yuan2011,Yue2019,Yuan2020,Niu2021}).

In the boxplots of $\nu_{2}-\nu_{1}$ in Figure \ref{fig:spectra_params}, some of the $\Delta \nu$ values inherit large uncertainties from $\nu_2$, especially for Ne, Si, Fe, Na, and Al.
Generally speaking, the $\nu_{2}-\nu_{1}$ values are the same for primary, secondary, and hybrid species within the uncertainties.
As the measurement of the spectral hardening, the $\Delta \nu$ of Si and Fe distribute around zero, which demonstrates the spectral hardening in these two species is not statistical significant. 
Moreover, it shows that the $\Delta \nu$ values of some primary CR nuclei species whose spectra have relative smaller uncertainties (proton, He, C, O, and Mg) are systematically smaller than that of the secondary species Li, Be, and B, which is the reason why some precious works claimed AMS-02 data (including the spectra or spectra ratio of Li, Be, B, C, and O) favors a break in diffusion coefficient index rather than a break in the primary source injection (see, e.g., \citet{Genolini2017,Niu2020}). 
What's more interesting is that the $\Delta \nu$ of F seems to be systematically smaller than that of Li, Be, and B. If we follow the conclusion obtained above (the spectral hardening of the secondary CR nuclei species dominately comes from the propagation process), it is an indication that the propagation properties of heavy cosmic rays, from F to Si, are different from those of light cosmic rays, from He to O \citep{AMS02_F}.

\section{Summary}

In summary, for the primary CR nuclei species, although $\nu_{1}$ and $\nu_{2}$ have similar values within uncertainties (except the $\nu_{1}$ of proton and Fe with special reasons), the significant different values of  $\Rbr$ indicate that their spectral hardening cannot come from an uniform mechanism in CR sources or propagation process.
A natural origin of the hardening is the superposition of different kinds of CR sources, which in the one hand can be corresponding to the galactic averaged CR sources and a local CR source (such as Geminga SNR \citep{Zhao2022} and the superflares from nearby M dwarfs \citep{Ohm2018}), and on the other hand can be correspond to different kinds of CR factories: such as the different population of supernova remnants \citep{Aharonian2004}, galactic center \citep{Scherer2022}, novas \citep{Nova_CR}, active red dwarf stars \citep{Sinitsyna2021}, etc.
  In both cases, as long as the CR sources have different elemental abundances, it will produce different $\nu_{2}-\nu_{1}$ and $\Rbr$ values for different CR nuclei species. The combination of the above two cases is also possible \citep{Zhang2022}.

For the secondary CR nuclei species, their concentrated values of $\Rbr$ are different from that of their parents species, which denies the possibility of the inheritance from the primary species and favors the propagation origin (such as the spatial dependent propagation \citep{Tomassetti2012,Guo2016}).
  Here, the different propagation regions can be corresponding to the structures of the galaxy (i.e., the galaxy center, the bulk, the disk, the halo, and even the spiral arms), in which the densities of ISM are different and thus they have different propagation environments.

As a result, the dominating factors of the spectral hardening for primary and secondary CR nuclei species are different. Of course, these factors will influence all the CR nuclei species spectra, regardless of the primary, the secondary or the hybrid ones, just with different weights.
The hybrid origins of the CR nuclei spectral hardening at a few hundred GV are also confirmed by \citet{Niu2022} via a propagation model.
  This hybrid origins will not only produce a break at about 200 GV in secondary/primary ratios (such as B/C and B/O), which corresponds to the dominating spectral hardening for secondary species; but also produce breaks greater than 200 GV in secondary/primary ratios, which corresponds to the dominating spectral hardening for primary species. These predictions are confirmed by the recently released B/C and B/O ratio from DAMPE \citep{DAMPE2022}. 
Moreover, the slightly different $\Delta \nu$ and $\Rbr$ distributions between the F and Li/Be/B show some hints that the propagation properties of heavy CRs are different from those of light CRs.


\section*{Acknowledgments}

Jia-Shu Niu would like to thank Hui-Fang Xue and Jue-Ran Niu for providing a quiet working environment.
This research was supported by the National Natural Science Foundation of China (NSFC) (No. 12005124 and No. 12147215).

\software{{\tt emcee} \citep{emcee}}


\end{CJK*}
\end{document}